\documentclass[twocolumn,showpacs,preprintnumbers,amsmath,amssymb,APSl,prd,nofootinbib,superscriptaddress]{revtex4-1}

\usepackage{bm}
\usepackage{mathrsfs}
\usepackage{xcolor,color,graphicx,graphics}
\usepackage[all]{xy}
\usepackage{epsfig,subfigure}
\usepackage{latexsym,amssymb,amsmath,amsfonts} 
\usepackage[english]{babel} 
\usepackage[OT1]{fontenc}
\usepackage[latin1]{inputenc}
\usepackage{makeidx}
\usepackage{hyperref}
\usepackage{color,graphicx,graphics,wrapfig,epsf}

\definecolor{red}{rgb}{1,0,0}

\def\+{^\dagger}

\def\<{\leftarrow}
\def\>{\rightarrow}

\def\({\left(}
\def\){\right)}

    \def\e{\epsilon}

\def\q{\quad}

\newcommand{\bi}{\begin{itemize}} 				\newcommand{\ei}{\end{itemize}}
\newcommand{\benu}{\begin{enumerate}} 		\newcommand{\enu}{\end{enumerate}}
\newcommand{\bd}{\begin{dinglist}{0}}     \newcommand{\ed}{\end{dinglist}}
\newcommand{\bfig}{\begin{figure}[htbp]}  \newcommand{\efig}{\end{figure}}
        			
\newcommand{\bc}{\begin{center}} 				  \newcommand{\ec}{\end{center}}
\newcommand{\be}{\begin{equation}} 				\newcommand{\ee}{\end{equation}}
\newcommand{\bsub}{\begin{subequations}}  \newcommand{\esub}{\end{subequations}}
\newcommand{\ben}{\begin{eqnarray}} 			\newcommand{\een}{\end{eqnarray}}
\newcommand{\ba}[1]{\begin{array}{#1}} 		\newcommand{\ea}{\end{array}}
\newcommand{\bea}{\begin{equation}\begin{array}{rcl}}
\newcommand{\eea}{\end{array}\end{equation}}

\begin{document}
\title{New scalar compact objects in Ricci-based gravity theories}

\author{Victor I. Afonso} \email{viafonso@df.ufcg.edu.br}
\affiliation{Unidade Acad\^{e}mica de F\'isica, Universidade Federal de Campina
Grande, 58.429-900 Campina Grande-PB, Brazil}
\affiliation{Departamento de F\'{i}sica Te\'{o}rica and IFIC, Centro Mixto Universidad de Valencia - CSIC. Universidad de Valencia, Burjassot-46100, Valencia, Spain}

\author{Gonzalo J. Olmo} \email{gonzalo.olmo@uv.es}
\affiliation{Departamento de F\'{i}sica Te\'{o}rica and IFIC, Centro Mixto Universidad de Valencia - CSIC.
Universidad de Valencia, Burjassot-46100, Valencia, Spain}
\affiliation{Departamento de F\'isica, Universidade Federal da
Para\'\i ba, 58051-900 Jo\~ao Pessoa, Para\'\i ba, Brazil}

\author{Emanuele Orazi} \email{orazi.emanuele@gmail.com}
\affiliation{ International Institute of Physics, Federal University of Rio Grande do Norte,
Campus Universit\'ario-Lagoa Nova, Natal-RN 59078-970, Brazil}
\affiliation{Escola de Ciencia e Tecnologia, Universidade Federal do Rio Grande do Norte, Caixa Postal 1524, Natal-RN 59078-970, Brazil}

\author{Diego Rubiera-Garcia} \email{drubiera@ucm.es}
\affiliation{Departamento de F\'isica Te\'orica and IPARCOS, Universidad Complutense de Madrid, E-28040 Madrid, Spain}

\date{\today}
\begin{abstract}

Taking advantage of a previously developed method, which allows to map solutions of General Relativity into a broad family of theories of gravity based on the Ricci tensor (Ricci-based gravities), we find new exact analytical scalar field solutions by mapping the free-field static, spherically symmetric solution of General Relativity (GR) into quadratic $f(R)$ gravity and the  Eddington-inspired Born-Infeld gravity.  The obtained solutions have some distinctive feature below the would-be Schwarzschild radius of a configuration with the same mass, though in this case no horizon is present. The compact objects found include  wormholes, compact balls, shells of energy with no interior, and a new kind of object which acts as a kind of wormhole membrane. The latter object has Euclidean topology but connects antipodal points of its surface by transferring particles and null rays across its interior in virtually zero affine time. We point out the relevance of these results regarding the existence of compact scalar field objects beyond General Relativity that may effectively act as black hole mimickers.

\end{abstract}

\maketitle

\section{Introduction} \label{sec:II}

In the last few decades, the investigation of gravitating solutions supported by scalar fields has blossomed with the finding of unexpected and striking results. Unfolded by such findings, many astrophysical applications have followed suit in the community. Among them we underline hairy black holes, namely, black holes with non-trivial scalar fields which seem to contradict the non-hair conjecture (see \cite{Herdeiro:2015waa} for a review), and new astrophysical compact configurations under the form of gravastars \cite{Visser:2003ge,Chirenti:2016hzd,Pani:2010em} and boson stars \cite{Colpi:1986ye,Palenzuela:2017kcg,Cunha:2017wao}, as well as other long-lived scalar field configurations \cite{Ayon-Beato:2015eca}. New phenomena can be triggered by such scalar fields and configurations, such as superradiance \cite{superreview} and black hole bombs \cite{Cardoso:2004nk,Sanchis-Gual:2015lje,Hod:2016kpm}.

Compact field objects have also been investigated within the context of modified theories of gravity \cite{Carballo-Rubio:2018jzw}. The latter are motivated by a blend of theoretical, astrophysical, and cosmological reasons, see e.g. \cite{DeFelice:2010aj, CLreview,Clifton:2011jh,Nojiri:2017ncd,Heisenberg:2018vsk} for some reviews on the subject. A fundamental difficulty when studying the phenomenology of such gravity theories resides in the (typically) highly non-linear nature of their field equations. This feature forces one to rely on approximate/perturbative methods and/or the use of specific ans\"atze. Also, the fact that existing numerical algorithms and codes are tightly attached to the structure of Einstein's field equations makes their implementation in modified theories of gravity very expensive from a computational point of view. These features largely prevent the extraction of significant theoretical signatures for any problem of physical interest. This is the case, for instance, of gravitational wave emission out of binary mergers (see e.g. \cite{Berti:2018vdi} for a discussion on this point). A very timely question within this context is the investigation of the existence of horizonless compact objects disguised as black holes, their observational signatures, and how to discriminate them
 \cite{Cardoso:2017cqb,Cardoso:2016rao}.

To overcome these difficulties, a novel approach to solve the field equations of modified theories of gravity formulated in metric-affine (or Palatini) spaces was recently introduced in \cite{Afonso:2018bpv} for general anisotropic fluids. The heart of this method lies on the existence of a duality between General Relativity (GR) coupled to some matter fields, described by a given Lagrangian density, and a family of theories built out of the (symmetrized) Ricci tensor and its contractions with the metric (dubbed Ricci-based gravity theories, or RBGs for short), coupled to the same kind of matter fields, but described by a different Lagrangian density. This way, the spaces of solutions of both theories can be mapped into each other via purely \emph{algebraic} transformations. This allows the full power of the analytical and numerical methods developed within the context of GR to play their magic on the RBG side. The reliability of this method has been verified for electric \cite{Afonso:2018mxn} and scalar \cite{Afonso:2018hyj} fields, where the hard-won solutions obtained by direct resolution of the RBG field equations (see for instance the cumbersome direct derivation of \cite{Afonso:2017aci} in the free scalar field matter case) were shown to be much easily re-obtained using this procedure.

The main aim of this work is to dig further into the applications of the method in the case of scalar field matter, by obtaining new analytical solutions. Since there might be more exotic compact objects than our imagination can grasp, generation of new solutions may indeed help us to parameterize the space of potential deviations from canonical solutions and their physical implications for future observations. In this work we shall anchor our derivations on the well known static spherically symmetric solution of GR coupled to a free real scalar field (with a canonical Lagrangian density), for which we refer to the work by Wyman in \cite{Wyman}. As target theories of the mapping on the RBG side we choose the ``Starobinsky" quadratic $f(R)$ gravity \cite{Staro}, and the Eddington-inspired Born-Infeld  (EiBI)  gravity originally proposed by Ba\~nados and Ferreira \cite{EiBI}, both of which have attracted a great deal of attention in the literature. Interestingly, the canonical dynamics of the scalar matter Lagrangian on the GR side gets mapped, on the RBG side, by somehow transferring the underlying non-linear structure of the RBG gravity Lagrangian density into the matter field sector. Namely, the quadratic $f(R)$ model gets coupled to a quadratic (non-canonical) scalar field, while the square-root structure of EiBI gravity ends up coupled to a similar square-root scalar field matter Lagrangian. The corresponding exact solutions in these two scenarios are thus directly obtained out of the GR scalar field solution \cite{Wyman} via the mapping.
We will discuss the main properties of the several new obtained configurations, some of which represent compact scalar field objects could be able to act as black hole mimickers.

This work is organized as follows: in Sec. \ref{sec:II} we introduce the Einstein-frame representation of the RBG field equations, briefly describe the mapping for scalar fields, and depict the main features of the GR free scalar field solution. The mapping of this solution for quadratic $f(R)$ gravity is described in Sec. \ref{sec:III} alongside the properties of the resulting solutions, and to EiBI gravity in Sec. \ref{sec:IV}. Finally, in Sec. \ref{sec:V} we summarize our results and discuss some future avenues of research.

\section{Implementing the map} \label{sec:II}

\subsection{Ricci-based gravities}

The RBG family of theories we shall consider in this work is built out as
\begin{equation} \label{eq:actionRBG}
\mathcal{S}=\frac{1}{2\kappa^2} \int d^4x  \sqrt{-g} \mathcal{L}_G(g_{\mu\nu}, R_{(\mu\nu)}(\Gamma)) + \mathcal{S}_{m}(g_{\mu\nu},\psi_m) \ ,
\end{equation}
with the following definitions and conventions: $\kappa^2$ is Newton's constant in suitable units, $g$ is the determinant of the space-time metric $g_{\mu\nu}$, which is independent of the affine connection $\Gamma \equiv \Gamma_{\mu\nu}$ (metric-affine or Palatini formalism); the connection alone allows to construct the (symmetrized) Ricci tensor $R_{(\mu\nu)}(\Gamma)=\partial_{\alpha}\Gamma^{\alpha}_{\nu\mu}-\partial_{\nu}\Gamma^{\alpha}_{\alpha\mu}+\Gamma^{\alpha}_{\alpha\beta}\Gamma^{\beta}_{ \nu \mu} -\Gamma^{\alpha}_{\nu\beta}\Gamma^{\beta}_{\alpha\mu}$ (from now on parenthesis will be removed), while the function $\mathcal{L}_G$ defines the RBG theory in terms of scalars built out as traces of the object ${M^\mu}_{\nu} \equiv g^{\mu\alpha}R_{\alpha\nu}$ and powers of it. Finally, $S_m=\int d^4x \sqrt{-g} \mathcal{L}_m(g_{\mu\nu},\psi_m)$ describes the matter sector, with $\psi_m$ denoting the matter fields. 

Two comments on the construction of this action are in order. First, torsion, the antisymmetric part of the connection, is assumed to vanish, which can be safely done as we are considering minimally coupled bosonic fields,
and the consequent projective invariance of the theory allows to remove torsion by a gauge choice (see \cite{Afonso:2017bxr} for details).
Second, as recently shown in Ref.\cite{BeltranJimenez:2019acz}, should one consider the antisymmetric part of the Ricci tensor in building the action (\ref{eq:actionRBG}), the theory would contain ghost-like instabilities, thus forcing us to deal with the symmetrized Ricci tensor only. Within these constraints, the RBG family contains, as particular cases, GR itself, $f(R)$, $f(R,R_{\mu\nu}R^{\mu\nu})$, EiBI gravity, and other extensions.

It has also been shown in \cite{Afonso:2017bxr} that the field equations derived from the action (\ref{eq:actionRBG}) admit an Einstein-frame representation of the form
\begin{equation} \label{eq:RBGeom}
{G^\mu}_{\nu}(q)=\frac{\kappa^2}{\vert \hat\Omega \vert^{1/2}} \left[ {T^\mu}_{\nu} -\left(\mathcal{L}_G +\frac{T}{2} \right) \delta_\mu^\nu \right] \ ,
\end{equation}
where $T{^\mu}_{\nu}\equiv \frac{-2}{\sqrt{-g}}\frac{\delta \mathcal{S}_m}{\delta g^{\mu\nu}}$ is the stress-energy tensor of the matter fields, and ${G^\mu}_{\nu}(q)$ is the Einstein tensor of an auxiliary metric $q_{\mu\nu}$, which is the one compatible with $\Gamma$, \emph{i.e.}, $\nabla_{\mu}^{\Gamma}(\sqrt{-q} q^{\alpha\beta})=0$ (that is, $\Gamma$ is Levi-Civita of $q$), and can be related to the space-time metric $g_{\mu\nu}$ via a deformation matrix $\hat\Omega$ as
\begin{equation} \label{eq:Omegadef}
q_{\mu\nu}=g_{\mu\alpha}{\Omega^\alpha}_{\nu} \ .
\end{equation}
This matrix $\hat\Omega$ (and the gravitational Lagrangian $\mathcal{L}_G$ too) can be written \emph{on-shell} as a function of the matter fields and (possibly) the space-time metric itself, being a model-dependent relation which can be explicitly found once a particular RBG theory is chosen. Explicit examples will be shown below. The field equations (\ref{eq:RBGeom}) feature their second-order and ghost-free character, recovering the GR solutions in vacuum, ${T^\mu}_{\nu}=0$, where they only propagate the two polarizations of the gravitational field (gravitational waves) travelling at the speed of light (see \cite{Jana:2017ost} for a recent discussion on this topic).

\subsection{Mapping GR and RBGs with a scalar field} \label{sec:IIB}

The representation (\ref{eq:RBGeom}) of the RBG field equations suggests that an identification with the Einstein equations of GR could be possible. This idea was fully implemented in Ref.\cite{Afonso:2018hyj} for different configurations involving an arbitrary number of scalar fields. The goal is to rewrite the right-hand side of (\ref{eq:RBGeom}) by consistently removing its dependence on $g_{\mu\nu}$ (which may appear via ${T^\mu}_{\nu}$, $\mathcal{L}_G$ and $\hat{\Omega}$) in favor of  $q_{\mu\nu}$ and the matter fields in such a way that it can be interpreted as the stress-energy tensor of another (nonlinear, in general) scalar field Lagrangian, namely, ${G^\mu}_{\nu}(q)=\kappa^2  \bar{T}{^\mu}_{\nu}$, where
\begin{equation} \label{eq:Tmunumapping}
\bar{T}{^\mu}_{\nu}\equiv q^{\mu\alpha}\bar{T}_{\alpha\nu}=
\frac{1}{\vert \hat \Omega \vert^{1/2}}\left[{T^\mu}_{\nu}-{\delta^\mu}_{\nu}\left(\mathcal{L}_G+\frac{T}{2}\right)\right] \ .
\end{equation}
This equation establishes a relation between the energy-momentum tensor on the RBG side, $T{^\mu}_{\nu}(g)$, and the one on the GR side, $\bar{T}{^\mu}_{\nu}(q)$, and can be explicitly formalized, besides scalar fields, for other matter sources such as (anisotropic) fluids \cite{Afonso:2018bpv}, and (non-linear) electromagnetic \cite{Afonso:2018mxn} fields.

Let us now briefly summarize the results of \cite{Afonso:2018hyj}  for the case of a single scalar field with vanishing potential, since this is the case we are interested in this work. In the RBG frame, the scalar field is described by an action
\begin{equation}
\mathcal{S}_m(X)=-\frac{1}{2} \int d^4x \sqrt{-g}P(X) \ ,
\end{equation} where $P(X)$ is some function of the kinetic invariant $X \equiv g^{\mu\nu} \partial_{\mu}\phi\partial_{\nu} \phi$, and whose stress-energy tensor reads ${T^\mu}_{\nu}= P_X {X^\mu}_{\nu}-(P/2) \delta^{\mu}_{\nu}$, where $P_X \equiv \partial P/\partial X$ and ${X^\mu}_{\nu} \equiv g^{\mu\alpha} \partial_{\alpha} \phi \partial_{\nu} \phi$ (so $X$ is simply its trace, $X={X^\mu}_{\mu}$). In the GR frame the scalar field will be described by a Lagrangian density $K(Z)$ minimally coupled to $q_{\mu\nu}$, \emph{i.e.}, with  $Z \equiv q^{\mu\nu} \partial_{\mu}\phi \partial_{\nu} \phi$ and with stress-energy tensor $\bar{T}{^\mu}_{\nu}= K_Z {Z^\mu}_{\nu}-(K/2) \delta^{\mu}_{\nu}$. If our RBG is $f(R)$, then the corresponding $K(Z)$ Lagrangian in the Einstein frame can be obtained from the RBG frame variables via (see \cite{Afonso:2018hyj} for details)
\begin{equation}\label{eq:direct_map}
K(Z)=\frac{1}{f_R^2}\left(\frac{f}{\kappa^2} +{X P_X-P}\right) \ .
\end{equation}
To use this expression in specific models, one must note that in $f(R)$ theories one has the relation $R f_R-2f=\kappa^2T$, which relates the scalar curvature $R$ with the trace of the matter fields, $T=XP_X-2P$, through some expression of the form $R=R(X)$. Given that in $f(R)$ theories one has ${\Omega^\mu}_{\nu}=f_R \delta^\mu_\nu$, then from (\ref{eq:Omegadef}) it follows the conformal relation  $g_{\mu\nu}=q_{\mu\nu}/f_R$ and, therefore, one has that $Z=X/f_R$, which can be used in (\ref{eq:direct_map}) to obtain (at least) a parametric representation of $K(Z(X))$.

The inverse problem, which is the one we are more interested in, involves going from GR coupled to $K(Z)$ to some RBG theory coupled to $P(X)$. In the $f(R)$ case, the Lagrangian $P(X)$ is given by
\begin{equation}\label{eq:inverse_map_f}
P(X)=\frac{f}{\kappa^2}+f_R^2\left(Z K_Z-K\right) \ .
\end{equation}
The relation between $Z$ and $X$ follows from the fact that $q^{\mu\nu}R_{\mu\nu}(q)=-\kappa^2\tilde{T}=-\kappa^2(ZK_Z-2K)=R/f_R$, which provides $R=R(Z)$, together with $X=Z f_R$. According to this, mapping GR coupled to $K(Z)=Z$ (canonical matter Lagrangian) into the Starobinsky $f(R)$ model (\ref{eq:fRquad}) leads to
\begin{equation} \label{eq:quadscalar}
P(X)=X+\alpha \kappa^2 X^2 \ ,
\end{equation}
where it is apparent that the explicit nonlinearity of the gravitational sector has been transferred to the matter Lagrangian.

In the case of EiBI gravity described by the action (\ref{eq:EiBIac}), the discussion of the general case is more involved because the relation between the metrics $g_{\mu\nu}$ and $q_{\mu\nu}$ is not conformal. Even so, one can obtain an explicit compact expression for the corresponding matter Lagrangian in the RBG frame when GR is coupled to $K(Z)=Z$, namely (check again \cite{Afonso:2018hyj} for details of this derivation)
\begin{equation}\label{eq:inverse_map_EiBI}
P(X)=\tfrac{2}{\epsilon \kappa^2} \left(\sqrt{1+\epsilon \kappa^2 X}-1\right) \ ,
\end{equation}
where we see how the structure of the gravitational side (here the square-root) is again transferred to the matter sector. This kind of Born-Infeld-type scalar field action has been used, for instance, within the context of accelerated cosmological solutions \cite{Fang:2004qj}.

\subsection{Static, spherically symmetric scalar field in GR}

In order to generate new solutions for RBGs we start with the well known analytic solution described by Wyman in \cite{Wyman}. This solution comes out of GR coupled to a massless scalar field with (canonical) Lagrangian density $K(Z)=Z$, and it describes the asymptotically flat space-time generated by a static, spherically symmetric scalar field configuration. The corresponding line element can be suitably written as
\begin{equation} \label{eq:linewy}
ds_{GR}^2=-e^{\nu}dt^2+\frac{e^\nu}{W^4}dy^2 +\frac{1}{W^2}(d\theta^2+\sin\theta^2d\varphi^2)  \ ,
\end{equation}
where $\nu$ and $W$ are functions of the radial coordinate $y$. The unusual form of this spherically symmetric line element is justified on the grounds that it leads to a very simple equation for the scalar field, namely, $\phi_{yy}=0$, which, indeed, allows for a direct resolution of the Einstein equations in that case. Without loss of generality, its solution can be taken as $\phi(y)=y$. Demanding asymptotic flatness, the metric functions in (\ref{eq:linewy}) take the form \cite{Wyman}
\begin{eqnarray}
e^\nu &=& e^{\beta y} \label{eq:nuWyman} \\
W &=& \gamma^{-1} e^{\beta y/2}\sinh(\gamma y) \label{eq:WWyman} \ ,
\end{eqnarray}
where the constant $\beta$ is related to the asymptotic Newtonian mass of the solution, $M$, as $\beta =-2M$, and we have the constant $\gamma\equiv \sqrt{\beta^2+2\kappa^2}/2$. It should be noted that in these coordinates the asymptotic limit corresponds to $y\to 0$, where the scalar field vanishes, while the center of the (spherical) solution is reached at $y \to \infty$.

In the next two sections we shall study the properties of the solutions that follow from the two non-canonical matter theories (\ref{eq:quadscalar}) and (\ref{eq:inverse_map_EiBI}) coupled to their respective RBGs ($f(R)$ and EiBI) starting with the GR canonical scalar field solution just described.

\section{$f(R)$ gravity} \label{sec:III}

In this section, our target RBG for implementing the mapping is the quadratic $f(R)$ model 
\begin{equation} \label{eq:fRquad}
f(R)=R-\alpha R^2\ ,
\end{equation}
which is also known as Starobinsky model \cite{Staro} (for an overall discussion of $f(R)$ theories in the Palatini formalism see \cite{Olmo:2011uz}). Here $\alpha$ is a constant with dimensions of length squared. As mentioned before, mapping a free scalar theory $K(Z)=Z$ from GR to this  gravity theory generates the quadratic scalar field Lagrangian (\ref{eq:quadscalar}). This kind of non-canonical kinetic scalar fields have been widely employed in the literature, for instance, as alternatives to dark energy \cite{ArmendarizPicon:2000ah} and inflation \cite{ArmendarizPicon:1999rj}, or as topological defects/solitonic models \cite{Bazeia:2007df,Babichev:2006cy,Adam:2007ag}.

Since for  $f(R)$ theories the relation between the Einstein frame metric $q_{\mu\nu}$ and the space-time metric $g_{\mu\nu}$ is simply conformal, one can use the relation $\kappa^2(2K-ZK_Z)=R/f_R$ to obtain $f_R=\frac{1}{1+2\alpha \kappa^2 Z}$, which leads to\footnote{In passing let us point out that this expression corrects the one presented in Eq.(71) of Ref.\cite{Afonso:2018hyj}, where the conformal factor $f_R$ wrongly appears in the numerator.}
\begin{eqnarray} \label{eq:f(R)le}
ds_{f(R)}^2=\frac{1}{f_R} ds^2_{GR}&=&(1+2\alpha \kappa^2Z)\Big[-e^{\nu}dt^2+\frac{e^\nu}{W^4}dy^2  \nonumber\\
&+&\frac{1}{W^2}(d\theta^2+\sin\theta^2d\varphi^2) \Big] \ ,
\end{eqnarray}
where $Z \equiv W^4 e^{-\nu}$, with the functions $\nu,W$ corresponding to those of Wyman's solution in Eqs.(\ref{eq:nuWyman}) and (\ref{eq:WWyman}), respectively. This line element provides a solution for the quadratic $f(R)$ gravity (\ref{eq:fRquad}) coupled to the quadratic free scalar field model (\ref{eq:quadscalar}), which can be seen as a deformation of the GR solution via the conformal factor $f_R$. This deformation has a very relevant physical impact on the features of the corresponding solutions, as we shall see next.

\subsection{Properties of the solution}

To start the physical analysis of the solutions, we first focus on the properties of the spherical sector in the line element (\ref{eq:f(R)le}), which is given by $r^2(y) \equiv g_{\theta \theta}$. In the asymptotically far region, that is $y \to 0$, this function behaves as $r^2(y) \approx 1/y^2 \rightarrow \infty$, exactly as in GR. However, near the center, which corresponds to large $y$, different behaviors arise according to the sign of $\alpha= s \vert \alpha \vert$, with $s =\pm 1$, which have nontrivial effects on the resulting geometry. We will study each case separately.

\subsubsection{$\alpha>0$}

\begin{figure}[t!]
 \begin{center}
\includegraphics[width=0.45\textwidth]{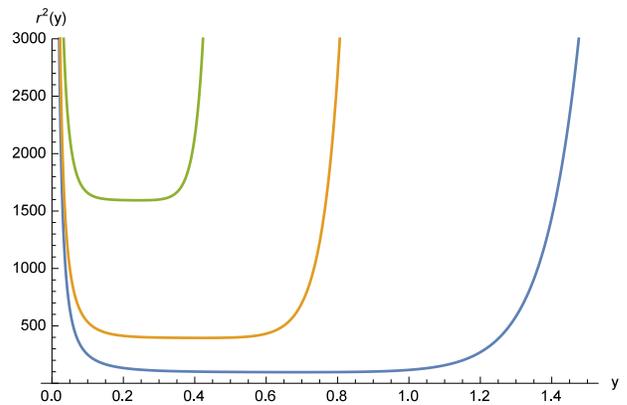}
\caption{Behaviour of the radial function $r^2(y)$ for quadratic $f(R)$ gravity (\ref{eq:fRquad}), in the $\alpha>0$ branch, where the  presence of the wormhole throat is clearly seen. In this figure we have set $\kappa^2=1$ and $\vert \alpha \vert=10^{-1}$, and chosen three values of $\beta=-10$ (blue), $\beta=-20$ (orange), and $\beta=-40$ (green). The location of the wormhole throat is only weakly varying with $\alpha$ but much more sensitive to $\beta$. In the limit $\alpha \to 0$ the wormhole throat closes and the radial function restores its GR behaviour, $r^2(y)\sim1/y^2$.}
\label{fig:whfr}
 \end{center}
\end{figure}

For $s=+1$, from Eq.(\ref{eq:f(R)le}) the $r^2(y)$ function takes the form
\begin{equation}
r^2(y)=\frac{1+2|\alpha|\kappa^2 e^{-\beta y}W^4}{W^2} \ ,
\end{equation}
which is always positive and well defined within the interval $y\in [0,\infty[$. Interestingly, since $W$ is a monotonically growing function and the term proportional to $|\alpha| $ is also multiplied by $e^{-\beta y}=e^{+2M y}>0$, which also grows as $y\to \infty$, it follows that the $|\alpha| $ correction will dominate at large $y$, implying the growth of the radial function $r^2(y)$ in this region. We thus see that $r^2(y)$ first decreases like $1/y^2$ as $y$ grows but then increases as it tends to infinity (see Fig. \ref{fig:whfr}), which puts forward the existence of a wormhole structure with its throat located at the minimum of $r^2(y)$ (for a broad account of wormhole physics see e.g. \cite{VisserBook}). The location of this minimum can be found analytically by taking $\sinh(\gamma y)\approx e^{\gamma y}/2$ in the region $y>1/\gamma$, which leads to
\begin{equation}\label{eq:ymin}
y_{min}\approx\frac{\log \left(\frac{\left(\beta^2+2 \kappa ^2\right)^{3/2} \left(\sqrt{\beta^2+2 \kappa ^2}+\beta \right)}{|\alpha|\kappa ^2 }\right)}{2 \sqrt{\beta^2+2 \kappa ^2}+\beta } \ .
\end{equation}
In the astrophysical limit, when $\vert \beta  \vert^2 \gg \kappa^2$, the above expression simplifies to
\begin{equation}
y_{min}\approx \frac{1}{\vert \beta \vert} \log \left(\frac{\beta^2 }{|\alpha| }\right) \ .
\end{equation}
Near that point we have
\begin{equation}
\lim_{y \to y_{min}}r^2(y) \approx \beta^2 + \left(3-\log\left[\frac{\beta^2}{\vert \alpha \vert }\right] \right)\kappa^2 + \mathcal{O}(\kappa^4) \ ,
\end{equation}
which implies that, for astrophysical configurations, the area of the wormhole throat  is smaller than the one of the event horizon of a Schwarzschild black hole ($r=2M$, with $\beta=-2M$ in our case) with the same mass. However, no event horizon is ever formed for these configurations.

To get deeper into the properties of the internal region of this structure ({\it i.e.} inside the throat), let us consider the geodesic motion of point-like particles. The geodesics for general static, spherically symmetric geometries with line element $ds^2=-C(y)dt^2 + B^{-1}(y)dy^2+r^2(y)d\Omega^2$, are described by the equation (see \cite{Olmo:2016tra} for details)
\begin{equation} \label{eq:geoeq}
\frac{C(y)}{B(y)}\left(\frac{dy}{d\lambda}\right)^2=E^2-C(y)\left(\frac{L^2}{r^2(y)}-k\right)
\end{equation}
where $\lambda$ is the affine parameter, $k=0,-1$ for null and time-like geodesics, respectively, and $E$ and $L$ are the energy per unit mass and angular momentum per unit mass, respectively. It is easy to see that the radial null geodesics ($k=0$, $L=0$) of this solution are the same as in GR because null rays are insensitive to the conformal factor. Since in GR those geodesics can get to $y\to \infty$ (the center of the object) in finite affine time, in our case it follows that the asymptotic region $\lim_{y\to\infty} r^2(y)\to\infty$ is also reached in a finite affine time. Since that region represents the asymptotic internal infinity, getting there in finite affine time means that the space-time is geodesically incomplete. Numerical integration of this equation for null non-radial and for time-like geodesics also leads to incomplete geodesics. A similar behavior was found in the context of EiBI gravity coupled to canonical scalar fields \cite{Afonso:2017aci}. This puts forward that though wormholes may sometimes help to indefinitely extend geodesics, their mere existence is not a guarantee for the geodesic completeness of the space-time.

\subsubsection{$\alpha<0$}

\begin{figure}[t!]
 \begin{center}
\includegraphics[width=0.45\textwidth]{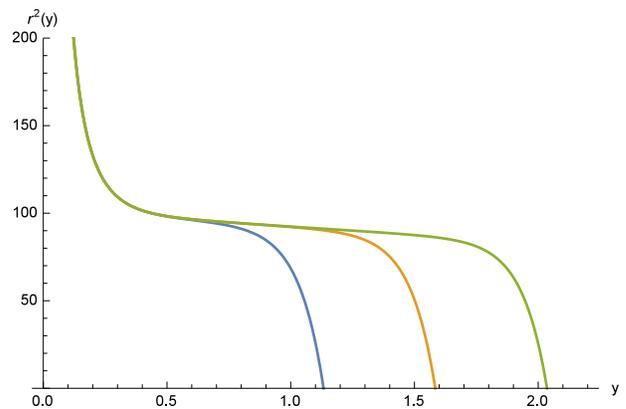}
\caption{Behaviour of the radial function $r^2(y)$ for quadratic $f(R)$ gravity (\ref{eq:fRquad}), in the $\alpha<0$ branch. In this figure we have set $\kappa^2=1$ and $\beta=-10$, and chosen three values of $\vert \alpha \vert=10^{-1}$ (blue),  $\vert \alpha \vert=10^{-3}$ (orange) and  $\vert \alpha \vert=10^{-5}$ (green). In the limit $\vert \alpha \vert \to 0$ the radial function restores its GR behavior.}
\label{fig:rmin}
 \end{center}
\end{figure}

\begin{figure}[t]
 \begin{center}
\includegraphics[width=8cm,height=5.1cm]{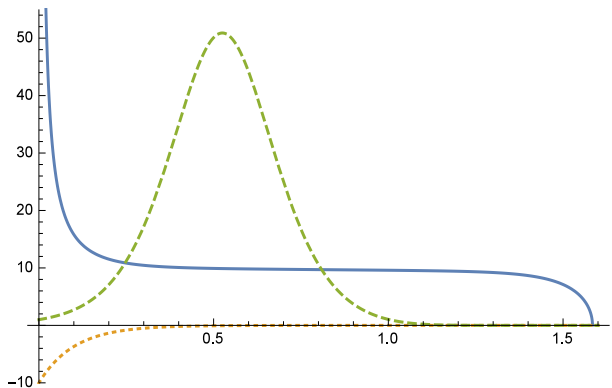}
\caption{Behaviour of the radial function $r^2(y)$ (solid blue), and the metric functions $10g_{tt}$ (dotted orange) and $g_{rr}$ (dashed green) for $\beta=-10$ and $\vert \alpha \vert=10^{-3}$. Note the large bump in the $g_{rr}$ component around the flattening region of $r^2(y)$.}
\label{fig:met}
 \end{center}
\end{figure}

For the branch $s=-1$ one finds that the radial function $r^2(y)$ crosses the region around the Schwarzschild radius $r=2M$ being nearly flat there but without reaching any minimum, after which it decreases sharply until reaching $r=0$ at a finite value of $y=y_{max}$ given by
\begin{equation}
y_{max}=\frac{\log \left(\frac{\left(\beta^2+2 \kappa ^2\right)^2}{\vert \alpha \vert }\right)}{2 \sqrt{\beta^2+2 \kappa ^2}+\beta}
\end{equation}
which marks the center of the configuration. The dependence of this radial function with $\vert \alpha \vert$ can be observed in Fig. \ref{fig:rmin}, where we point out that this  effect becomes more acute with growing $\vert \beta \vert$.
To get further into the interpretation of this structure let us consider the behaviour of the metric functions. As depicted in Fig. \ref{fig:met} for a typical sample with $\beta=-10$, while the $g_{tt}$ metric component shows a monotonic behavior starting at asymptotic flatness ($g_{tt} \to -1$ at $y=0$) towards the center of the solution, the metric component $g_{rr}$ (also asymptotically flat, $g_{rr} \to 1$ at $y=0$) has a huge but bounded bump at the flattening region of $r^2(y)$, before relaxing to zero as the maximum value of the field is reached (corresponding to $r=0$ and $y=y_{max}$).  Note that since the amplitude of the bump is bounded, no event horizon is formed in this case either.  This object represents essentially a compact ball of scalar matter whose area is very approximately the same as expected in a Schwarzschild black hole of the same mass. On the other hand, its geodesic structure for null radial geodesics is exactly the same as the one of the positive branch case and the one of GR, since photons do not see the conformal factor.

\subsection{Curvature and energy density}

\begin{figure}[t!]
 \begin{center}
\includegraphics[width=8cm,height=5.1cm]{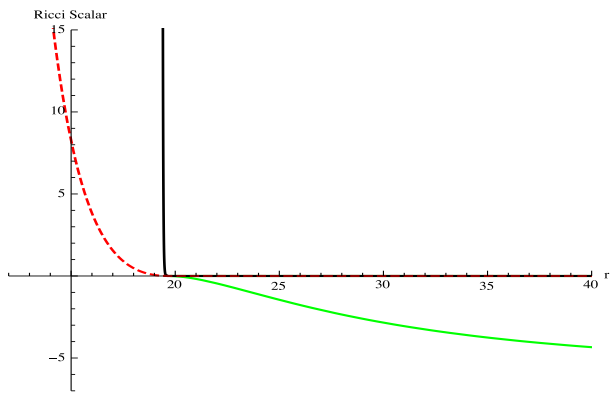}
\caption{Ricci scalar as a function of  $r(y)$ in the case of GR (black solid), $\alpha>0$ (solid green), and $\alpha<0$ (dashed red). All three curves converge outside the body and are indistinguishable. The green curve on the negative vertical axis represents the curvature of the $\alpha>0$ case in the interior region of the wormhole. }
\label{fig:Ricc}
 \end{center}
\end{figure}

In order to explore the interior geometry of the above configurations in more detail, it is useful to approximate the hyperbolic functions in the metric by exponentials, as we did above to get Eq.(\ref{eq:ymin}). This allows us to obtain compact expressions for quantities such as the Ricci scalar or the effective energy density in regions of interest. One can numerically verify that such approximations are extremely accurate in all relevant cases. Proceeding in that way, one finds that the scalar curvature near the center in the GR solution diverges exponentially fast as
\begin{equation}\label{eq:R_GR}
\lim_{y\to\infty} R_{GR}\approx \frac{\kappa ^2 e^{ \left(2 \sqrt{\beta ^2+2 \kappa ^2}+\beta \right)y}}{\left(\beta ^2+2 \kappa ^2\right)^2} \ .
\end{equation}
When the quadratic $f(R)$ corrections are taken into account, the growth of the Ricci scalar is considerably softened. Focusing on the astrophysical limit, $|\beta|^2 \gg \kappa^2$, we find (see Fig. \ref{fig:Ricc})
\begin{eqnarray}
\lim_{y\to\infty} R_{\alpha>0}&\approx & - \frac{3 \beta ^2}{2 \alpha } \\
\lim_{y\to y_{max}} R_{\alpha<0}&\approx & -\frac{3}{2 |\alpha|  \beta  \left(y-y_{max}\right)^3} \ .
\end{eqnarray}
The divergence of the above curvature scalars in the $\alpha\leq 0$ branch is also accompanied by a divergent energy density there. In the regular (wormhole) case ($\alpha >0$), however, the inner geometry quickly tends to a de Sitter plus  dust effective stress-energy tensor. This  follows from evaluation of the Einstein tensor with the corresponding line element, which leads to
\begin{equation}
{G^\mu}_\nu= \left(
\begin{array}{cccc}
 \frac{3\beta ^2}{4 \alpha } & 0 & 0 & 0 \\
 0 & \frac{\beta ^2}{4 \alpha } & 0 & 0 \\
 0 & 0 & \frac{\beta ^2}{4 \alpha } & 0 \\
 0 & 0 & 0 & \frac{\beta ^2}{4 \alpha } \\
\end{array}
\right) = \frac{\beta ^2}{4 \alpha } \delta^\mu_\nu+\rho_{dust}u^\mu u_\nu    \ ,
\end{equation}
where the fluid energy density is $\rho_{dust}=-\frac{\beta ^2}{2 \alpha }<0$. It is worth noting that though the wormhole case has finite energy density and finite curvature everywhere, the internal region is geodesically incomplete because radial null geodesics reach the $r\to \infty$ region in finite affine time, as shown in the previous section. In the other (non-wormhole) cases, radial null geodesics always hit a curvature divergence, which is usually interpreted as a pathology. A more detailed analysis of their impact on physical observers is required to draw any conclusions because the intensity of those divergences is quite different.

\section{EiBI gravity} \label{sec:IV}

Let us now consider the EiBI theory as target of our analysis. Its action can be conveniently written as (see \cite{BeltranJimenez:2017doy} for a recent review on this theory, its extensions and phenomenology)
\begin{equation} \label{eq:EiBIac}
\mathcal{S}_{EiBI}=\frac{1}{\e\kappa^2} \int d^4x\left[\sqrt{\vert g_{\mu\nu} + \epsilon R_{\mu\nu}\vert} - \lambda \sqrt{-g}\right] \ .
\end{equation}
In this theory, the (length-squared) parameter $\epsilon$ determines the scale at which deviations from GR become non-negligible; in this sense, EiBI gravity reduces to GR at curvature scales $\vert R_{\mu\nu} \vert \ll 1/\epsilon $. Moreover, the theory features an effective cosmological constant given by $\Lambda=(\lambda-1)/\kappa^2$. Hereafter we shall focus on asymptotically flat solutions, $\lambda=1$.

As discussed in Sec. \ref{sec:IIB}, for EiBI gravity  a canonical free scalar field model in GR maps into the non-canonical Lagrangian density of Eq.(\ref{eq:inverse_map_EiBI}). Therefore, similarly to what happens in the $f(R)$ case, the non-linear (square-root) structure of EiBI gravity ends up transferred to the scalar field sector. As opposed to the $f(R)$ case, though, for EiBI gravity the relation (\ref{eq:Omegadef}) between metrics is not conformal, but given by a deformation matrix ${[\Omega^{-1}]^{ \mu}}_\nu$, which in the case of free scalar fields takes the form
\begin{eqnarray}
\label{eq:Omega-1BIY}
{[\Omega^{-1}]^{ \mu}}_{\!\nu}\!&=& \tilde{A}\, {\delta^\mu}_{\nu}+ \tilde{B} {Z^\mu}_\nu \\
\tilde A\, &\equiv&\tilde A(Z)= 1-\tfrac{\epsilon\kappa^2}{2}(K-Z K_Z) \q \\
\tilde B\,&\equiv&\tilde B(Z)= -\epsilon\kappa^2K_Z \ ,
\end{eqnarray}
Since we are dealing with $K(Z)=Z$ on the GR side, the above deformation matrix becomes simply
\begin{equation}\label{eq:Omega-1map}
{[\Omega^{-1}]^{ \mu}}_\nu=  {\delta^\mu}_{\nu}-\epsilon\kappa^2 {Z^\mu}_\nu \ .
\end{equation}
As a result, from Eq.(\ref{eq:Omegadef}) we find that
\begin{equation}\label{eq:Omega-1map}
g_{ \mu\nu}= q_{\mu\nu}-\epsilon\kappa^2 Z_{\mu\nu} \ ,
\end{equation}
which leads to the form of the EiBI line element
\begin{eqnarray}
ds_{EiBI}^2&=&-e^{\nu}dt^2+\left(\frac{e^\nu}{W^4}-\epsilon \kappa^2\right)dy^2 \nonumber  \\
&+&\frac{1}{W^2}(d\theta^2+\sin\theta^2d\varphi^2)  \ , \label{eq:BImapping}
\end{eqnarray}
where we have again used that $\phi(y)=y$. Note the simplicity of this line element as compared to that corresponding to GR, see Eq.(\ref{eq:linewy}). The only difference is a constant shift $\to \epsilon\kappa^2$ in the $g_{yy}$ component of the metric.

\subsection{Properties of the solution}

Likewise in the $f(R)$ case, here we also have two different branches of solutions depending on the sign of $\epsilon$. However, in this case, as readily seen from (\ref{eq:BImapping}), the spherical sector and the $g_{tt}$ part of the metric do not change their functional dependences on the radial function in this case as compared to GR, being $g_{yy}$ the only difference. Recall that from  (\ref{eq:BImapping}) the radial function can be written as
\begin{equation}
r^2(y)=\frac{1}{W^2(y)}=\tfrac{1}{4} e^{-\beta y}(\beta^2+2\kappa^2)\,\text{csch}^2\left[\tfrac{1}{2}\sqrt{\beta^2+2\kappa^2} \right]\ .
\end{equation}
which always yields a monotonic behavior for $r^2(y)$ such that  $r^2(y) \simeq 1/y^2 $ in the far limit ($y \to 0$) and $r^2(y)\approx \left(\beta ^2+2 \kappa ^2\right) e^{-\left(\sqrt{\beta ^2+2 \kappa ^2}+\beta \right)y}$ as the center is approached in the limit $y\to \infty$.

\subsubsection{$\epsilon >0$}

In this case the $g_{yy}$ component vanishes always at a finite value of $y$ while $g_{tt}$ remains negative. As a result, since signature changes should not be allowed, the physically acceptable region should be restricted to the interval in which $g_{yy}\ge 0$. For astrophysical configurations, $ \vert \beta \vert^2 \gg \kappa^2$, this value can be explicitly found to be
\begin{equation} \label{eq:ycEiBIp}
y_c \approx \frac{\log \left(\frac{\left(\beta ^2+2 \kappa ^2\right)^2}{\kappa ^2 |\epsilon| }\right)}{2 \sqrt{\beta ^2+2 \kappa ^2}+\beta }\approx  \frac{\log \left(\frac{\beta ^4}{\kappa ^2 |\epsilon| }\right)}{|\beta| }\ .
\end{equation}
Expansion of the radial function $r^2(y)$ around this point and considering the limit $\beta^2\gg \kappa^2$ leads to
\begin{equation}
r_c^2(y_c) \approx \beta^2 -\left(\log\left[\frac{\beta^4}{\vert \epsilon  \vert \kappa^2}\right]-2\right)\kappa^2 + \mathcal{O}(\kappa^4)\ .
\end{equation}
Since $|\beta|=2M$, one readily verifies that this value is always smaller than the Schwarzschild radius corresponding to that same mass configuration. This result implies that this new object is remarkably different from its GR counterpart. In fact, the GR solution represents a compact ball of scalar energy that extends from $r\lesssim 2M$ down to $r=0$. In this solution of the EiBI theory, the object has essentially the same external properties as in GR but has no interior region below $r=r_c$. This unusual object is also plagued with undesired divergences. Indeed, a careful look at the Ricci scalar shows that it diverges at $y_c$ as $R\sim 1/(\epsilon\kappa^2(y-y_c)^2)$, which is of polynomial type. The components of the corresponding Einstein tensor also diverge at that point, which is equivalent to having unbounded effective energy density and pressures. Such divergences are also found in the GR solution, though they appear at $y\to \infty$ (or $r\to 0$) and grow exponentially with $y$. One can verify that radial null geodesics reach $y=y_c$ after a finite affine time which, together with the impossibility of extending the geometry beyond that surface, implies that the space-time is geodesically incomplete. There seems to be no more remarkable features of these configurations.

\begin{figure}[t!]
 \begin{center}
\includegraphics[width=0.454\textwidth]{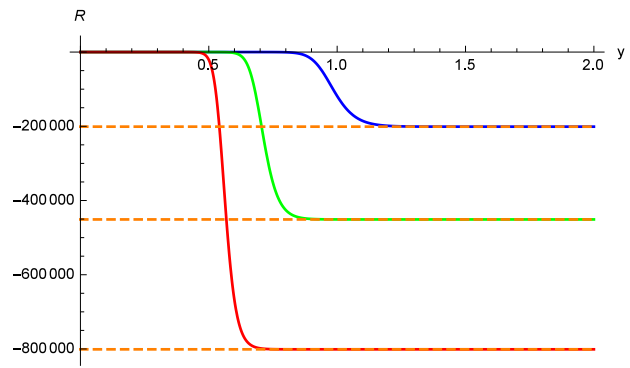}
\caption{Behavior of the Ricci scalar $R \equiv g^{\mu\nu}R_{\mu\nu}(g)$ of EiBI gravity for $\epsilon<0$. In this figure we have set $\kappa^2=1$, $\epsilon=-10^{-3}$ and three values of $\beta=-20$ (blue), $\beta=-30$ (green) and $\beta=-40$ (red). In dashed orange we represent the asymptotic values of the Ricci scalar in the limit $y \to \infty$ (central region) for each case, as given by Eq.(\ref{eq:Ricciasym}).}
\label{fig:Ricci}
 \end{center}
\end{figure}

\subsubsection{$\epsilon <0$}

\begin{figure}[t!]
 \begin{center}
\includegraphics[width=8cm,height=6.8cm]{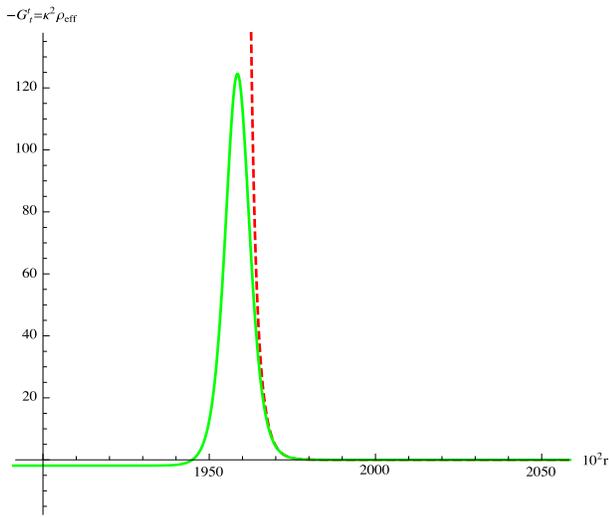}
\caption{Representation of the effective energy density $\kappa^2\rho_{eff}=-{G^t}_t$ of the EiBI solutions. The dashed (red) curve corresponds to $\epsilon>0$, which diverges, while the solid (green) curve is the $\epsilon<0$ case, which is well-behaved everywhere. Here the object has a mass $M=10$ (in units of $\kappa^2$). Note that the peak is located below the corresponding Schwarzschild radius, which here would correspond to $10^2r=2000$.}
\label{fig:EiBI_density}
 \end{center}
\end{figure}

When $\epsilon$ is negative the metric component $g_{yy}$ is everywhere nonvanishing and positive, and the solutions are defined all the way down to the center ($y \to \infty$). A glance at the line element  puts forward that for large values of $y$ the function $g_{yy}\approx |\epsilon|\kappa^2+\left(\beta ^2+2 \kappa ^2\right)^2e^{-|\beta| y}$ rapidly tends to a constant in such a way that the {\it proper radial distance} to the center is infinite:
\begin{equation}\label{eq:properdistance}
\lim_{y\to\infty} L=\int^y \sqrt{g_{yy}}dy\sim \sqrt{|\epsilon|\kappa^2}y\to \infty \ .
\end{equation}
 In the inner region below $r<|\beta|=2M$, the scalar curvature quickly tends to a constant value (see Fig.\ref{fig:Ricci}), which for astrophysical sources ($|\beta|^2 \gg \kappa^2$) can be approximated by (see Fig. \ref{fig:Ricci})
\begin{equation} \label{eq:Ricciasym}
R(y \to \infty) \to -\frac{\beta ^2}{2 |\epsilon| \kappa^2 } .
\end{equation}
The region that defines the transition from outside to inside of the object has additional interesting features. A glance at the exact expression of the ${G^t}_t$ component of the Einstein tensor, which defines an effective energy density for the matter that generates this geometry as $ \kappa^2\rho_{eff}=-{G^t}_t$, shows that the effective energy density is finite everywhere and concentrated on a bump or thick shell peaked at $r\lessapprox |\beta|$, as shown in Fig. \ref{fig:EiBI_density}. The interior region then turns into a ball of constant, negative energy density for which
the corresponding Einstein tensor also becomes constant. Thus, apparently, no pathologies seem to be present in this solution. However, a more careful analysis of the limit towards the center brings up a residual contribution which cannot be neglected as $y\to \infty$. In fact, one finds that both the Einstein tensor and the scalar curvature  have divergent contributions in that limit. The complete expression for the Einstein tensor must take into account the contributions from $\kappa^2$, and becomes

\onecolumngrid
\begin{equation}
{G^\mu}_\nu=\left(
\begin{array}{cccc}
 \frac{3 \left(\beta ^2+\beta\sqrt{\beta ^2+2 \kappa ^2}  +\kappa ^2\right)}{2 |\epsilon| \kappa^2 }-\frac{e^{\left(\beta +\sqrt{\beta ^2+2 \kappa ^2}\right) y}}{\beta ^2+2 \kappa ^2} & 0 & 0 & 0 \\
 0 & -\frac{e^{\left(\beta +\sqrt{\beta ^2+2 \kappa ^2}\right) y}}{\beta ^2+2 \kappa ^2} & 0 & 0 \\
 0 & 0 & \frac{2 \beta ^2+\sqrt{\beta ^2+2 \kappa ^2} \beta +2 \kappa ^2}{4 |\epsilon| \kappa^2 } & 0 \\
 0 & 0 & 0 & \frac{2 \beta ^2+\sqrt{\beta ^2+2 \kappa ^2} \beta +2 \kappa ^2}{4 |\epsilon| \kappa^2 } \\
\end{array}
\right)
\end{equation}
\twocolumngrid

Analogously, the complete expression for the scalar curvature reads
\begin{equation}\label{eq:Ricci_EiBI}
R=-\frac{5 \beta ^2+4 \beta  \sqrt{\beta ^2+2 \kappa ^2}+5 \kappa ^2}{2 |\epsilon| \kappa^2 }+\frac{2 e^{ \left(\sqrt{\beta ^2+2 \kappa ^2}+\beta \right)y}}{\beta ^2+2 \kappa ^2} \ ,
\end{equation}
which recovers (\ref{eq:Ricciasym}) in the $\kappa\to 0$ limit. It is worth noting that this quantity diverges as $R\sim e^{\kappa^2y/|\beta|}$ but with substantially less intensity than the GR expression (\ref{eq:R_GR}), $R\sim e^{|\beta| y}$. Thus, for astrophysical sources this divergence is strongly suppressed  in the EiBI theory while  it is strongly magnified in GR.

\begin{figure}
 \begin{center}
\includegraphics[width=8cm,height=6.1cm]{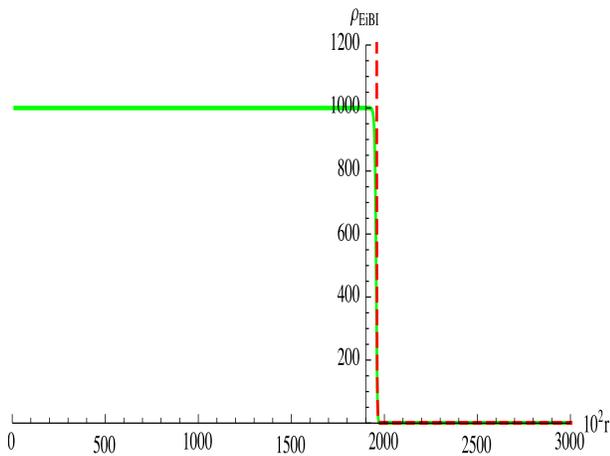}
\caption{Representation of the energy density $\rho_{EiBI}=-{T^t}_t$ for the scalar field Lagrangian (\ref{eq:inverse_map_EiBI}) when $\epsilon<0$ (solid, green) and when $\epsilon>0$ (dashed, red). In the $\epsilon<0$  case, a transition from the vacuum value $\rho_{EiBI}=0$ to its internal maximum value $\rho_{EiBI}=1/(|\epsilon|\kappa^2)$ takes place below the corresponding Schwarzschild radius $r\lessapprox |\beta|$, here located at $10^2r=2000$. In the $\epsilon>0$  case, the canonical energy density diverges at that point.}
\label{fig:kink}
 \end{center}
\end{figure}

To deepen into the physical implications of the above divergences, it is useful to consider some basic notions on the energy associated to these objects. If one reads the energy density directly from the stress-energy tensor of the scalar field, then in the GR solution we have $\rho_{GR}\equiv -{T^t}_t=Z/2=q^{yy}/2$, which in the internal region goes like $\rho_{GR} \sim e^{|\beta| y}$. This divergence in the matter profile is automatically transferred to the geometric sector via Einstein's equations, which justifies the blow up of (\ref{eq:R_GR}).  In the EiBI case, instead, we have $\rho_{EiBI}=P(X)/2$, which in the internal region rapidly tends to $\rho_{EiBI}\approx  1/(|\epsilon|\kappa^2)$, which is a positive constant that saturates the natural scale of the Born-Infeld Lagrangian (\ref{eq:inverse_map_EiBI}). A complete representation of $\rho_{EiBI}$ appears in Fig. \ref{fig:kink}, where the scalar field energy density adopts a kink-like profile. Thus, by measuring the scalar field and its energy density, we would conclude that this solution is physically acceptable. The picture drawn by interpreting the Einstein tensor as an observable effective stress-energy tensor provides a completely different physical picture that has nothing to do with the actual behavior of the matter field. This suggests that the divergences of ${G^\mu}_\nu$ and the Ricci scalar (\ref{eq:Ricci_EiBI}) can be regarded as artifacts with little or no physical meaning.

\begin{figure}
 \begin{center}
\includegraphics[width=0.45\textwidth]{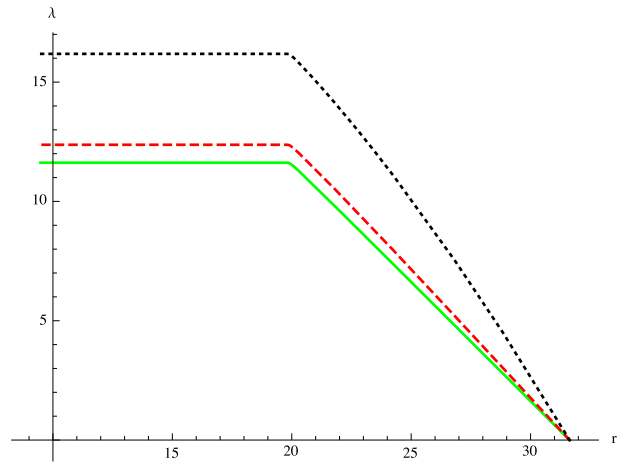}
\caption{Ingoing geodesics corresponding to EiBI with $\epsilon<0$. Upon crossing the surface of the compact object, around $r\lessapprox 20$, light cones get dramatically modified, allowing for an almost instantaneous transfer of light rays and massive particles from the surface to the center and from there to the exterior again. In a sense, this works like a wormhole but with Euclidean topology. The solid (green) line represents a radial null geodesic, dashed (red) is null with angular momentum, and the dotted (black) is a timelike geodesic with angular momentum.  }
\label{fig:geodesics_EiBI}
 \end{center}
\end{figure}

To complete our analysis, let us now have a look at the geodesic structure of the solution. For null radial geodesics, the line element (\ref{eq:BImapping}), together with the condition of energy conservation, $E=e^{\nu}dt/d\lambda$, lead to
\begin{equation}
\left(\frac{d(E\lambda)}{dy}\right)^2={e^\nu}\left({\frac{e^\nu}{W^4}+|\epsilon|\kappa^2}\right) \ .
\end{equation}
In the far limit, $y\to 0$, one has $e^\nu\approx 1$ and $W=1/r\approx y$, which leads to $(dr/d\lambda)^2=1$ and represents light rays propagating at the speed of light ($c=1$), $r(t)=r_0\pm t$, as expected. The opposite limit corresponds to $y\to \infty$, and in this case we have $(d(E\lambda)/dy)^2\approx e^{-|\beta| y}|\epsilon|\kappa^2$, which leads to $E\Delta \lambda=\mp(2\sqrt{|\epsilon|\kappa^2}/|\beta|)e^{-|\beta|y/2}$. This shows that the effective speed of light rays inside the object, $dy/d\lambda$, grows exponentially fast near the center, $y\to \infty$ (see Fig. \ref{fig:geodesics_EiBI}). As a result, a ray that crosses the surface of the object takes a ridiculously small fraction of affine time $\Delta \lambda \propto e^{-|\beta|y_c/2}$ to get to the center and then to go from the center to its antipodal point, whith $y_c$ corresponding to the surface of the object.   A similar behavior takes place for non-radial rays and massive particles, whose geodesic equation degenerates into that of radial null geodesics as soon as the interior of the object is reached.
At this stage, it is worth noting that despite the fact that the proper radial distance  from the surface to the center computed using (\ref{eq:properdistance}) is infinite, information and particles can cross the object in virtually zero time. Had we defined the physical proper radial distance from the surface to the center in operational terms, i.e., as $\Delta L=c \Delta \lambda$, where $c$ is the speed of light and $\Delta \lambda$ the proper time computed above for a signal that goes from the surface to the center of the object, then we would conclude that  $\Delta L\to 0$. Thus, in practical terms, it is as if the interior did not exist, with this object acting like a wormhole that transfers information and particles from one point on the surface to its antipodal point almost instantaneously. This  contrasts with the case $\epsilon>0$, for which the geometry below the surface is not well defined due to a change of signature of the metric that would lead to two time-like coordinates, and because of the blowup of both the canonical and effective energy densities at the surface (see Fig. \ref{fig:EiBI_density} and Fig. \ref{fig:kink}).

\section{Conclusion and perspectives} \label{sec:V}

In this work we have worked out and discussed exact analytical solutions for compact scalar field objects in two well motivated extensions of GR, namely, quadratic $f(R)$ and Eddington-inspired Born-Infeld gravity, formulated in metric-affine spaces. These two models are particular examples of a large family of gravitational theories (dubbed as Ricci-based gravities) constructed in terms of scalars out of the Ricci tensor, and which admit an Einstein-frame representation of their field equations. This representation allows to find new solutions using a correspondence with the GR field equations whose main elements were discussed in detail, for scalar fields, in \cite{Afonso:2018hyj}.

Starting with a well known static, spherically symmetric solution of a free canonical scalar field in GR, we have generated new exact solutions in the above mentioned gravity theories coupled to (non-canonical) scalar Lagrangians. The correspondence presented in \cite{Afonso:2018hyj} is what justifies the choice of such non-canonical Lagrangians. We have found that the resulting objects typically show, for astrophysical configurations, some distinctive features close to (but below) the would-be Schwarzschild radius of a black hole with the same mass, though in our case no horizon is ever present. Indeed, depending on the theory and the branch of solutions (defined by the sign of the characteristic gravitational parameter of each model), we have found evidence of wormhole structures, compact balls, shells with no interior, and an object which admits a double interpretation depending on how one looks at the energy density that generates the geometry. In this latter case, which in our opinion is the most interesting one, the canonical stress-energy tensor describes a localized object whose radius is smaller than the corresponding Schwarzschild radius, and which has vanishing density outside $r\approx 2M$ and a rapid transient to a constant maximum positive density inside. On the other hand, if one interprets the Einstein tensor of this solution as an effective stress-energy tensor, the object looks like a thin shell supported by a negative pressure interior. Attending to its observational features, this object seems to act as a wormhole membrane which transfers particles and light from a point on the surface to its antipodal point in vanishing time.

The results obtained in this work for scalar fields contrast significantly with those found previously for electric fields in RBGs in many respects. Though electrically charged sources also generate wormholes and objects without interior (see e.g. \cite{Bambi:2015zch}), their effects on the background geometry are typically restricted to the innermost parts of the geometry, far from the would-be event horizon, thus having little to none impact on astrophysical phenomena. In particular, the structure of external horizons is generically preserved, while the innermost part develops new model-dependent properties \cite{Menchon:2017qed}. This is so because the modified gravitational dynamics of RBGs is strongly dependent on the local stress-energy densities. For electrovacuum solutions, the energy density is concentrated near the center, while in the case of scalar fields it begins to grow significantly already near the Schwarzschild radius. This triggers the modified gravity effects already at macroscopic scales and, as a result, it can have a nontrivial impact on astrophysically relevant scales, giving rise to new forms of black hole mimickers. In this sense, though orbital motions around the  solutions studied here would be essentially identical to those of a Schwarzschild black hole, their corresponding shadows are expected to be quite different due to the lack of an event horizon \cite{Cunha:2015yba}.

The bottom line of this analysis is that modified gravity theories which are thought to be relevant only at Planckian scales may also have a significant phenomenological impact on astrophysical scales. This, together with the possibility of extending the solutions of GR to other gravity theories, motivates a systematic exploration of new solutions in order to broaden the spectrum of known compact objects and identify peculiar observational features that may help discriminate among them from astrophysical data. An analysis of the stability and perturbative properties of such objects seems thus necessary in order to confront their properties with gravitational wave data (see, for instance, the analysis of  \cite{Cardoso:2016oxy,Konoplya:2019nzp} for potential smoking guns in this context). Further work along these lines is currently underway. \\

\section*{Acknowledgments}

 GJO is funded by the Ramon y Cajal contract RYC-2013-13019 (Spain). DRG is funded by the \emph{Atracci\'on de Talento Investigador} programme of the Comunidad de Madrid (Spain) No. 2018-T1/TIC-10431, and acknowledges further support from the Funda\c{c}\~ao para a Ci\^encia e
a Tecnologia (FCT, Portugal) research projects Nos. PTDC/FIS-OUT/29048/2017 and PTDC/FIS-PAR/31938/2017.  VIA is partially supported by Federal University of Campina Grande, Brazil. This work is supported by the Spanish projects FIS2014-57387-C3-1-P and FIS2017-84440-C2-1-P (MINECO/FEDER, EU), the project H2020-MSCA-RISE-2017 Grant FunFiCO-777740, the project SEJI/2017/042 (Generalitat Valenciana), the Consolider Program CPANPHY-1205388, the Severo Ochoa grant SEV-2014-0398 (Spain) and the Edital 006/2018 PRONEX (FAPESQ-PB/CNPQ, Brazil).  This article is based upon work from COST Action CA15117, supported by COST (European Cooperation in Science and Technology). VIA and DRG thank the Department of Physics and IFIC of the University of Valencia for their hospitality during different stages of the elaboration of this work. We finally thank Fernando Barbero for useful discussions.

\end{document}